\documentclass[fleqn,a4paper,english,11pt]{article}
\usepackage{fullpage,babel,amsmath,amsfonts,amsgen,amssymb,amsbsy,graphicx,physics,xcolor,flushend,longtable}
\usepackage[numbers,square,sort&compress]{natbib}
\usepackage[hidelinks,colorlinks=true,citecolor=blue,linkcolor=blue,urlcolor=blue]{hyperref}

\newcommand{\eqn}[2]{\begin{gather}
#1
\label{#2}
\end{gather}
}

\title{\bf Shape uncertainty analysis of laminar forced convection in a round microchannel\\ with viscous dissipation}
\author{{\bf A. Barletta}$^{*,1}$, {\bf M. Celli}$^{1}$, {\bf L. A. Sphaier}$^{2}$,\\[4pt] {\bf P. V. Brand\~ao}$^{1}$, {\bf S. Lazzari}$^{3}$, {\bf E. Ghedini}$^{1}$}

\date{\small
$^1$ Department of Industrial Engineering, Alma Mater Studiorum Universit\`a di Bologna,\\ Viale Risorgimento 2, 40136 Bologna, Italy\\[4pt]
$^2$ Department of Mechanical Engineering, Universidade Federal Fluminense,\\ Rua Passo da Pátria 156, sala 302, bloco D, Niterói, RJ, Brazil\\[4pt]
$^3$ Department of Architecture and Design, Università degli Studi di Genova,\\ Stradone di S.Agostino 37, 16128 Genova, Italy
}
\begin{document}

\maketitle

\begin{abstract}\noindent
The shape of a microchannel is usually affected by a significant uncertainty due to the small size of the cross-section, comparable with the typical wall-roughness length scale. Such an uncertainty is present at any scale, but it is definitely amplified and, hence, significantly important when the hydraulic diameter becomes smaller than some tenth micrometers. The focus of this paper is on the numerical analysis of the sensitivity to shape random modification, within a prefixed maximum extent, of the main flow and heat transfer characteristics of fully-developed forced convection. The numerical solutions are carried out by employing a finite-element solver repeated over a statistical sample of randomly generated perimetral profiles that simulate the wall roughness. The study includes the evaluation of the Fanning friction factor and of the Nusselt number for the {\sf T}, {\sf H1} and {\sf H2} thermal boundary conditions. \\[12pt]
\end{abstract}

\subsection*{Nomenclature}
\begin{longtable}[l]{ll}
$\langle \cdot\, \rangle$ & average value over the statistical sample\\
$a$ & constant temperature gradient\\
$Br$ & Brinkman number\\
$c$ & specific heat\\
$f$ & Fanning friction factor\\
$k$ & thermal conductivity\\
$\vb n$ & unit normal vector\\
$Nu$ & Nusselt number\\
$p$ & pressure\\
$\cal P$ & cross-sectional perimeter\\
$q_w$ & average wall heat flux\\
$r$ & radial coordinate\\
$r_0$ & duct radius\\
$Re$ & Reynolds number\\
$\cal S$ & cross-sectional area\\
$T$ & temperature\\
$T_b$ & bulk temperature\\
$T_w$ & average wall temperature\\
$u$ & velocity\\
$u_0$ & reference velocity\\
$u_m$ & mean flow velocity\\
$x,y,z$ & Cartesian coordinates\\[6pt]
{\em Greek symbols} \\
$\alpha$ & thermal diffusivity\\
$\Delta(\cdot)$ & standard deviation\\
$\Phi$ & dissipation function\\
$\mu$ & dynamic viscosity\\
$\nu$ & kinematic viscosity\\
$\sigma$ & Poiseuille number, $f\,Re$\\[6pt]
 {\em Subscripts, superscripts}\hspace{-1.5cm} \\
$\bar{~}$ & dimensionless quantity\\
{\sf T}, {\sf H1}, {\sf H2} & thermal boundary conditions\\
\end{longtable}
\setcounter{table}{0}

\section{Introduction}
Forced convection flows in microchannels are of great interest for the applications in the field of thermal management \cite{siddiqui2017efficient}. For instance, an important application is the regenerative cooling of combustion chamber liners, where large heat flux removal at diverse operation temperatures. Furthermore, studies of microfluidics and microscale heat transfer have a big impact on the cryogenic cooling employed for the supercomputing chips. Cooling of concentrator photovoltaics via microchannel fluid flows turned out to be an innovative application, as pointed out by \citet{gilmore2018microchannel}.

Miniaturization of the heat exchangers for several energy and heat transfer applications often entails the use of microchannels. A significant uncertainty affects the shape of a microchannel cross-section due to the small hydraulic diameter. When the hydraulic diameter becomes smaller than some tenth micrometers, the shape uncertainty induced by the wall roughness, and present at any scale, becomes significantly important. The focus of this paper is on the numerical analysis of the sensitivity to shape random modifications, within a prefixed maximum extent, of the main flow and heat transfer characteristics of fully-developed forced convection with viscous dissipation. The study includes the evaluation of the Fanning friction factor and of the Nusselt number for the {\sf T}, {\sf H1} and {\sf H2} thermal boundary conditions.

There is a wide literature regarding the effect of viscous dissipation on the forced convection analysis of microchannels. On the other hand, there is little information regarding the effects of random departures from the average shape of the cross-section. Such departures are expected to be important due to the roughness of the solid wall, which depends on the manufacturing techniques and on the surface treatment of the materials. Their effects lead to uncertainties in the significant parameters characteristic of the forced convection flow and heat transfer. The originality of this study is the numerical character of the analysis carried out by a finite-element solver on a domain where the boundary is randomly modified with respect to the circular shape.

We point out that considering a random geometry as a model of wall roughness appears to be an interesting approach. Many authors model the wall roughness via regular geometric patterns, as in the study by \citet{wang2007influence} where the wall roughness is represented through periodic wall functions. A similar approach can also be found in \citet{turgay2009effect} and in \citet{croce2007three} where regular distributions of conical peaks have been envisaged. Random generation of peaks and valleys of rectangular shape was carried out in the computational study by \citet{croce2005numerical}. We finally note that experimental data assessing the role of wall roughness were reported by \citet{celata2002experimental} relative to the transition from laminar to turbulent regimes.

\section{Governing equations}
Let us consider a straight round microchannel and allow for an uncertainty in the actual shape of the cross-section, due to the practical manufacturing characteristics whatever is the adopted fabrication technique. 

Shape uncertainty issues may be due to the size of the typical surface roughness relative to the small hydraulic diameter of the microchannel. Thus, though the average shape of the microchannel is circular with radius $r_0$, we will denote the perimeter of the duct cross-section with $\cal P$ and encircled area with $\cal S$. More precisely, $\cal P$ is a closed polygon which matches a circle within a tolerance $\gamma$, with $0 < \gamma < 1$, meaning that each point on $\cal P$ has a distance $r$ from the centre of the cross-section such that $(1-\gamma)\,r_0 < r < (1+\gamma)\,r_0$. Hereafter, we will set $\gamma = 1/10$. We note that the symbols $\cal P$ and $\cal S$ will be tacitly employed to denote both the geometric objects and their measures (perimeter and area).

Let $(x,y)$ be the Cartesian coordinates on the cross-sectional plane and $z$ be the streamwise coordinate. 
The flow is assumed to be laminar, stationary, incompressible and fully-developed, so that the only nonzero component of the velocity field is that in the $z$ direction, which is denoted with $u$. Due to the local mass balance equation, {\em i.e.} the zero-divergence condition on the velocity, $u$ is independent of $z$, and hence $u=u(x,y)$. Furthermore, the local difference between the pressure and the hydrostatic pressure depends only on $z$ and is denoted as $p(z)$. Under such conditions, the local momentum balance equation is expressed as
\eqn{
\laplacian{u} = \frac{1}{\mu}\, \dv{p}{z},
}{1}
where $\dd p/\dd z$ is a constant and $\mu$ denotes the dynamic viscosity. Thus, \eqref{1} is a two-dimensional partial differential equation defined in the $(x,y)$ plane. Such an equation is subject to the no-slip boundary condition
\eqn{
u = 0 \qc (x,y) \in {\cal P}.
}{2}
%
The thermal boundary conditions envisaged in this study are either {\sf T}, {\sf H1} or {\sf H2} according to the classification defined by \citet{shah1978laminar}. We recall that condition {\sf T} means a peripherally (along ${\cal P}$) and longitudinally (along $z$) uniform wall temperature, {\sf H1} means a peripherally uniform wall temperature and a longitudinally uniform wall heat flux, while {\sf H2} means a peripherally and longitudinally uniform wall heat flux. It is also well-known that conditions {\sf H1} and {\sf H2} coincide when ${\cal P}$ is exactly a circle \cite{shah1978laminar}. Furthermore, the boundary conditions {\sf T}, {\sf H1} and {\sf H2} imply that the derivative $\partial T/\partial z$ is a constant \cite{shah1978laminar, BARLETTA199615}, hereafter denoted as $a$.
Thus, the local energy balance equation is given by \cite{shah1978laminar, BARLETTA199615},
\eqn{
\alpha \, \laplacian{T} - a\, u + \frac{\nu}{c}\, \Phi = 0 \qc \Phi = | \grad{u} |^2 ,
}{3}
where $\alpha$, $\nu$ and $c$ are fluid properties, namely, the thermal diffusivity, the kinematic viscosity and the specific heat. Moreover, $\Phi=\Phi(x,y)$ is the viscous dissipation function \cite{BARLETTA199615}. The importance of viscous dissipation effect for the local energy balance in a microchannel has been demonstrated by many authors \cite{KOO20043159, BXu2003, van2009effect, Nonino2010, jing2017joule, mukherjee2017effects}.

\subsection{Dimensionless formulation}
Let us define the dimensionless quantities and operators
\eqn{
(\bar{x}, \bar{y}) = \frac{(x,y)}{r_0} \qc \bar{\grad} = r_0 \grad \qc \bar{\nabla}^2 = r_0^2 \nabla^2 \qc \bar{u} = \frac{u}{u_0} \qc \bar{T} = k\,\frac{T - T_w}{q_w\, r_0},
}{4}
where $k$ is the thermal conductivity of the fluid and
\eqn{
u_0 = - \frac{2 r_0^2}{\mu}\, \dv{p}{z} \qc T_w = \frac{1}{\cal P} \int\limits_{\cal P} T \, \dd{\cal P} \qc  q_w = \frac{k}{\cal P} \int\limits_{\cal P} \vb{n}\vdot\grad{T} \, \dd{\cal P}.
}{5}
Here, $q_w$ is the peripherally averaged incoming wall heat flux, while $u_0$ is a constant reference velocity such that
\eqn{
\frac{u_0}{u_m} = f \, Re = \sigma \qc u_m = \frac{1}{{\cal S}} \int\limits_{\cal S} u \, \dd {\cal S},
}{6}
with $u_m$ the mean flow velocity, $f$ the Fanning friction factor and $Re = 2\, u_m\, r_0/\nu$ the Reynolds number \cite{shah1978laminar}. We note that the mean wall temperature $T_w$ is a constant, {\em i.e.}, it is independent of $z$ only when the {\sf T} boundary condition is selected. With either conditions {\sf T}, {\sf H1} or {\sf H2}, it is easily proved that the mean wall heat flux $q_w$ is a constant and the constant $a$ not only coincides with $\partial T/ \partial z$, but also with $\dd T_w/\dd z$ and $\dd T_b/\dd z$, where $T_b$ is the bulk temperature,
\eqn{
\pdv{T}{z} = \dv{T_w}{z} = \dv{T_b}{z} = a  \qc T_b = \frac{1}{{\cal S}\, u_m} \int\limits_{\cal S} T u \, \dd {\cal S} . 
}{7}
In fact, the bulk value of a scalar is nothing but its mean value over the cross-section ${\cal S}$ weighted by the velocity profile $u(x,y)$.

On account of \eqref{5} and \eqref{7}, the dimensionless temperature $\bar T$ defined by \eqref{4} has interesting properties as
\eqn{
\int\limits_{\cal P} \bar{T} \, \dd{\cal P} = 0 \qc \dv{\bar{T}}{z} = 0 \qc \frac{1}{{\cal S}\, u_m} \int\limits_{\cal S} \bar{T} u \, \dd {\cal S} = - \frac{2}{Nu} ,
}{8}
where the Nusselt number $Nu$ is a constant whose definition is given by \cite{shah1978laminar},
\eqn{
Nu = \frac{2 r_0\, q_w}{k \qty(T_w - T_b)} .
}{9}
On account of \eqref{4}, one can rewrite \eqref{1} as
\eqn{
\bar{\nabla}^2 \bar{u} + \frac{1}{2} = 0 ,
}{10}
to be solved by employing the boundary condition
\eqn{
\bar{u} = 0 \qc (\bar{x},\bar{y}) \in \bar{\cal P}.
}{11}
The symbols $\bar{\cal P}$ and $\bar{\cal S}$ denote the dimensionless geometric objects obtained from ${\cal P}$ and ${\cal S}$ with the coordinates scaled by the reference length $r_0$. 
After solving \eqref{10} and \eqref{11}, one can evaluate the Poiseuille number $\sigma = f\, Re$ by employing \eqref{5} and \eqref{6} through the formula
\eqn{
\sigma = f\, Re = \frac{\bar{\cal S}}{\displaystyle\int\limits_{\bar{\cal S}} \bar{u} \, \dd \bar{\cal S}}.
}{13}
On account of \eqref{4}, the local energy balance equation \eqref{3} can be rewritten as
\eqn{
\bar{\nabla}^2\bar{T} - \sigma\, \bar{a}\, \bar{u} + 2\, \sigma^2 Br\, | \bar{\grad}{\bar{u}} |^2 = 0 ,
}{14}
where $Br$ is the Brinkman number and $\bar{a}$ is a dimensionless parameter,
\eqn{
Br = \frac{\mu u_m^2}{2 r_0 q_w} \qc \bar{a} = \frac{k r_0 u_m}{q_w \alpha}\, a .
}{15}
By employing Gauss' theorem together with the definitions given by \eqref{4} and \eqref{5}, an integration of \eqref{14} over the dimensionless region $\bar{\cal S}$ yields an explicit expression of $\bar{a}$, namely
\eqn{
\bar{a} = \frac{\bar{\cal P}}{\bar{\cal S}^{\vphantom{0^2}}} + \frac{2 \,\sigma^2 Br}{\bar{\cal S}^{\vphantom{0^2}}} \int\limits_{\bar{\cal S}} | \bar{\grad}{\bar{u}} |^2 \, \dd \bar{\cal S} .
}{16}
The boundary conditions to be satisfied are, alternatively,
\eqn{
\bar{T} = 0 \qc (\bar{x},\bar{y}) \in \bar{\cal P} \qc \text{for the boundary condition {\sf T}},
}{17}
\eqn{
\bar{T} = 0 \qc (\bar{x},\bar{y}) \in \bar{\cal P} \qc \text{for the boundary condition {\sf H1}},
}{18}
\eqn{
\vb{n}\vdot\bar{\grad}\bar{T} = 1  \qc (\bar{x},\bar{y}) \in \bar{\cal P} \qc \int\limits_{\bar{\cal P}} \bar{T} \, \dd\bar{\cal P} = 0 \qc \text{for the boundary condition {\sf H2}},
}{19}
The steps to be followed for every assignment of the cross-sectional region $\bar{\cal S}$ are the following:
\begin{enumerate}
\item Solve \eqref{10} and \eqref{11} for the dimensionless velocity profile $\bar{u}\qty(\bar{x}, \bar{y})$;
\item Evaluate $\sigma$ by using \eqref{13};
\item The local energy balance equation \eqref{14} must be solved for the dimensionless temperature profile $\bar{T}\qty(\bar{x}, \bar{y})$ with a boundary condition given either by \eqref{17}, \eqref{18} or \eqref{19}. 
\end{enumerate}

\subsubsection{Boundary condition \textsf{\textbf{T}}}
In the case {\sf T}, the Brinkman number is unknown a priori, while $\bar{a}$ must be set equal to $0$. Thus, \eqref{16} serves to determine the value of $Br$ to be employed in the solution of \eqref{14} and \eqref{17},
\eqn{
Br = - \frac{\bar{\cal P}}{\displaystyle 2 \,\sigma^2 \int\limits_{\bar{\cal S}} | \bar{\grad}{\bar{u}} |^2 \, \dd \bar{\cal S}} .
}{20}

\subsubsection{Boundary conditions \textsf{\textbf{H1}} and \textsf{\textbf{H2}}}
In the cases {\sf H1} and  {\sf H2}, $Br$ is an input parameter with any arbitrary value and $\bar{a}$ is to be evaluated by employing \eqref{16} prior to the solution of \eqref{14} with either conditions \eqref{18} or \eqref{19}. It is important to recall that in the special case where the uncertainty in the exact shape of the duct cross-section is zero, namely the case where $\bar{\cal S}$ is the unit disk, the two boundary conditions {\sf H1} and  {\sf H2} coincide \cite{shah1978laminar}. This is a consequence of the special geometrical symmetry  of the circle, which is broken by the surface roughness of the duct.

\subsection{Evaluation of the Nusselt number}
After taking the above described steps toward the determination of $\bar{u}$ and $\bar{T}$, one can finally evaluate $Nu$ by using \eqref{6}-\eqref{8}, namely
\eqn{
Nu = - \frac{2\, \bar{\cal S}}{\displaystyle \sigma \int\limits_{\bar{\cal S}} \bar{T} \, \bar{u} \, \dd \bar{\cal S}} .
}{21}

\begin{figure}[t]
\centering
\includegraphics[width=0.45\textwidth]{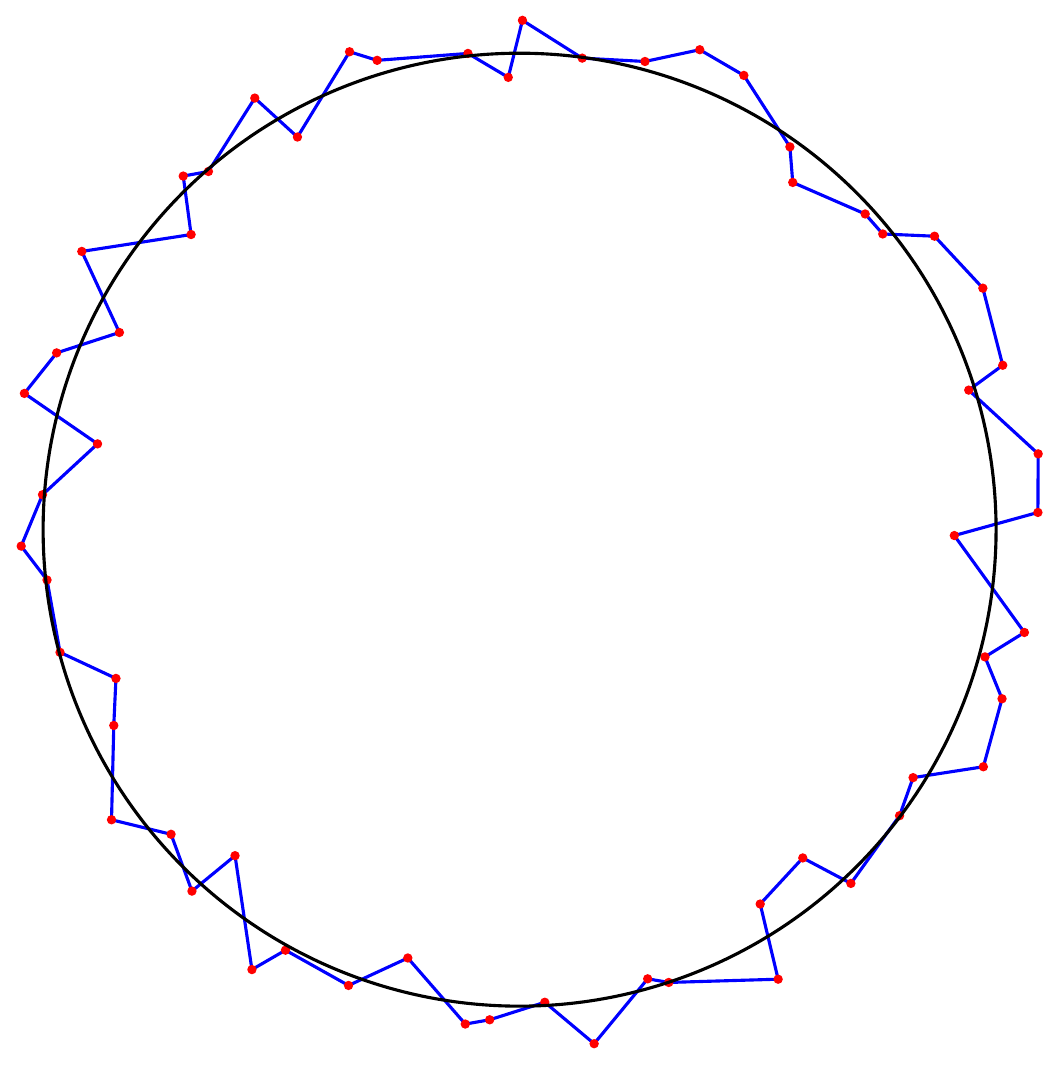}(a) \quad
\includegraphics[width=0.45\textwidth]{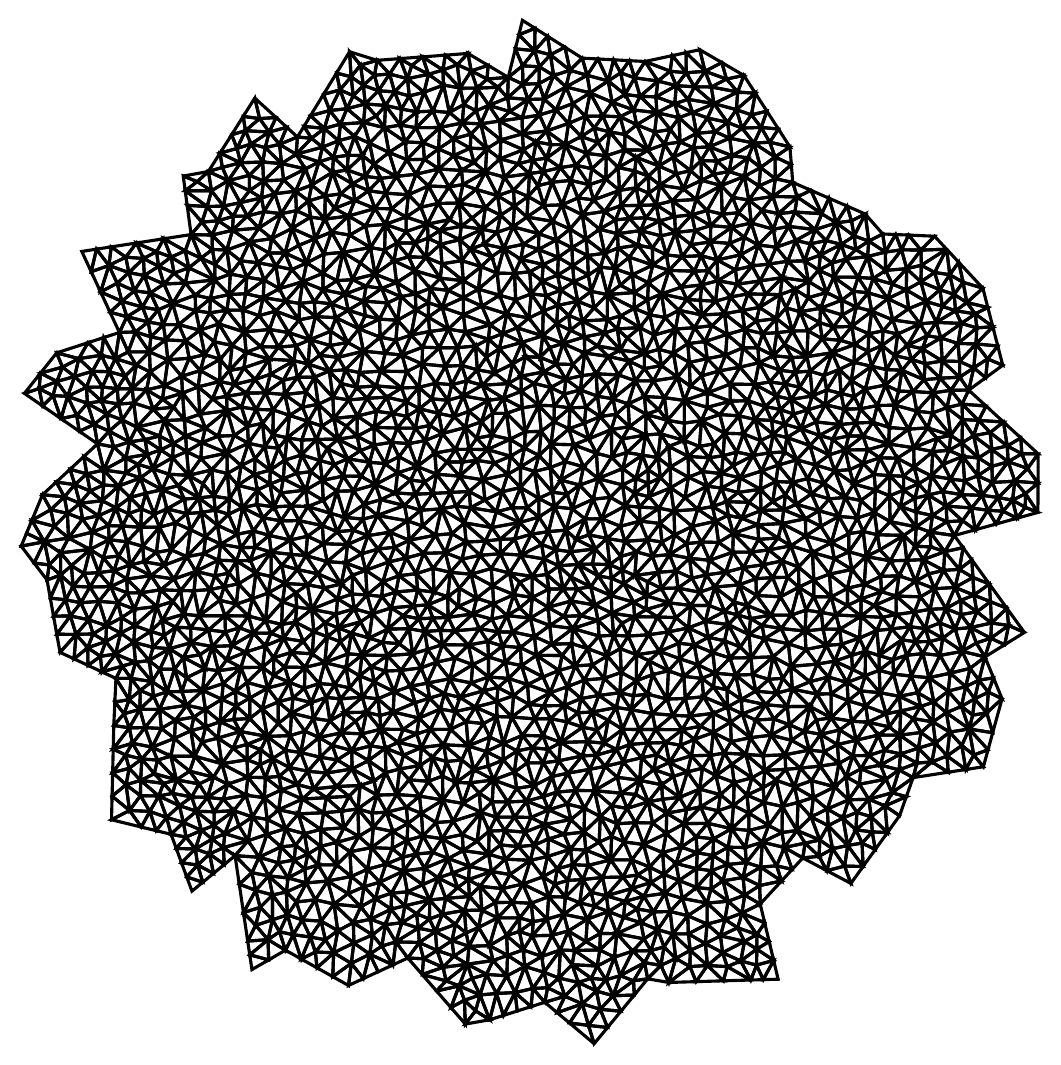}(b)
\caption{\label{fig1}\rm\small (a) Sample polygonal approximation of the actual microchannel boundary (blue) as compared to the smooth circular boundary (black); (b) Unstructured mesh for the actual computational domain}
\end{figure}

\begin{table}[t]
\centering
\begin{tabular}{|c|c|c|c|}
\hline
{\tt MaxCellMeasure} & $\sigma$ & $Nu$ & $Br$\\
\hline
0.01       & 17.1320 & 7.57290 & -0.154092\\
\hline
0.005     & 17.0886 & 7.53763 & -0.154483\\
\hline
0.001     & 17.0254 & 7.48563 & -0.155057\\
\hline
0.0005   & 17.0140 & 7.47562 & -0.155161\\
\hline
0.0001   & 17.0001 & 7.46315 & -0.155288\\
\hline
0.00005 & 16.9975 & 7.46079 & -0.155311\\
\hline
\end{tabular}
\caption{\label{tab1}\rm\small Accuracy test of the numerical solution}
\end{table}

\subsection{A perfectly smooth round microchannel}
Recalling the well-known results relative to an exactly circular cross-section may be useful for the forthcoming discussion of the results. In the absence of any roughness, the dimensionless velocity is given by the Hagen-Poiseuille profile,
\eqn{
\bar{u} = \frac{1}{8} \qty(1 - \bar{r}^2) ,
}{22}
with $\bar{r} = \sqrt{\bar{x}^2 + \bar{y}^2}$.
Moreover,
\eqn{
\bar{\cal S} = \pi \qc \bar{\cal P} = 2\pi . 
}{23}
Thus, \eqref{13} and \eqref{16} yield
\eqn{
\sigma = f\, Re = 16 \qc \bar{a} = 2 \qty( 1 + 8\, Br) . 
}{24}
One can now employ \eqref{14}, \eqref{17}-\eqref{19} and \eqref{21}. Then, for the boundary condition {\sf T}, one has
\eqn{
Br = - \frac{1}{8} \qc \bar{T} = - \frac{1}{4} \qty(1 - \bar{r}^4) \qc Nu = \frac{48}{5} .
}{25}
for the boundary condition {\sf H1} and {\sf H2}, one has
\eqn{
\bar{T} = - \frac{1}{4} \qty(1 - \bar{r}^2) \qty[3 - \bar{r}^2 + 16 Br \qty(1 - \bar{r}^2)]  \qc Nu = \frac{48}{11+48 Br} .
}{26}

\begin{figure}[t]
\centering
\includegraphics[width=0.35\textheight]{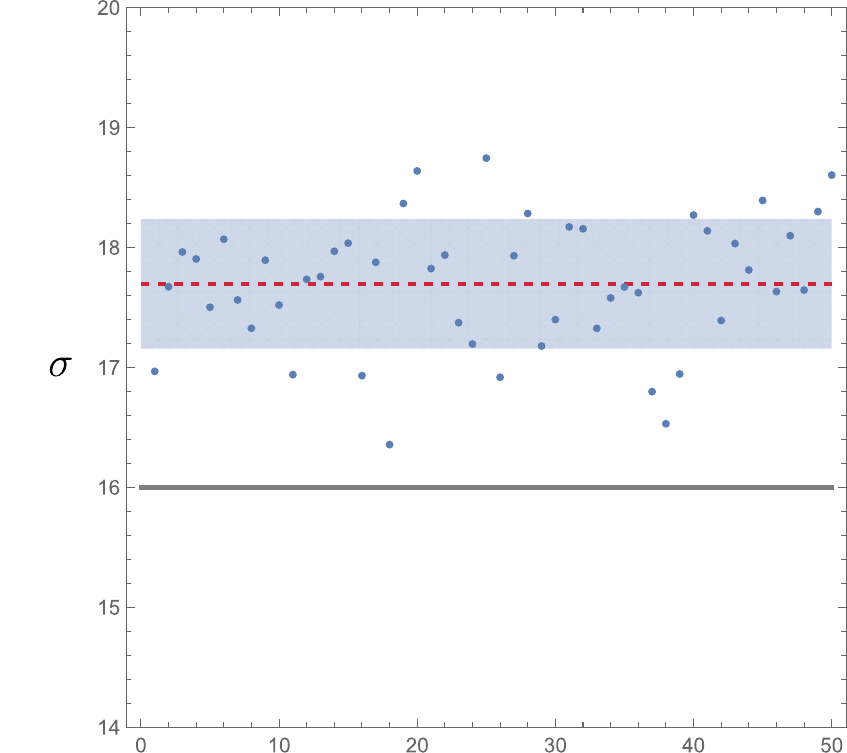}
\caption{\label{fig2}\rm\small Values of $\sigma$ (blue dots) for the 50 different regions; the red dashed line indicates the average value $\langle\sigma\rangle$ and the blue band denotes the range $\langle\sigma\rangle\pm\Delta\sigma$, where $\Delta\sigma$ is the standard deviation; the grey line is for $\sigma=16$ relative to the smooth circular microchannel}
\end{figure}

\begin{figure}[t]
\centering
\includegraphics[width=0.35\textheight]{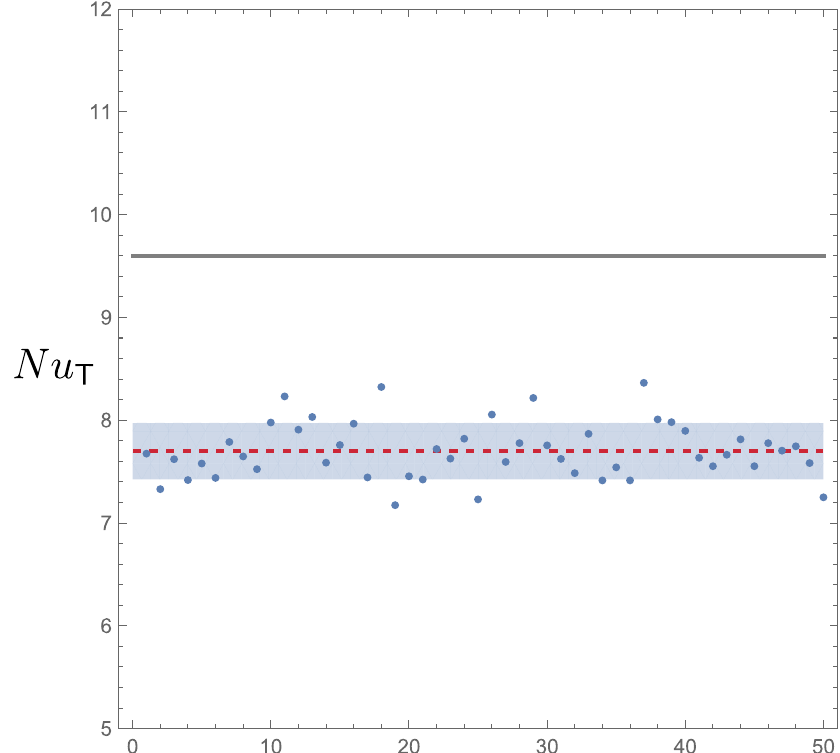}
\caption{\label{fig3}\rm\small Values of $Nu$ with the {\sf T} boundary condition (blue dots) for the 50 different regions; the red dashed line indicates the average value $\langle Nu \rangle$ and the blue band denotes the range $\langle Nu\rangle\pm\Delta Nu$, where $\Delta Nu$ is the standard deviation; the grey line is for $Nu=48/5$ relative to the smooth circular microchannel}
\end{figure}

\section{Numerical methodology}
The path defining the dimensionless boundary $\bar{\cal P}$ of the microchannel cross-section $\bar{\cal S}$  is defined by a polygon (see Fig.~\ref{fig1}). The points connected by the polygonal path are randomly generated with polar coordinates by assigning angles between $0$ and $2\pi$ and radiuses within the range $[0.9, 1.1]$. In Fig.~\ref{fig1}, the number of points used for generating the polygon is $60$. The computational domain is the dimensionless region bounded by the polygonal path. Uniform unstructured meshes are generated with triangular elements. The domain and mesh generation is achieved by using the software {\sl Mathematica 14} (\copyright{~}Wolfram Research Inc.). Furthermore, the weak formulation of the governing equations and their solution with the finite element method is managed by utilising the built-in function {\tt NDSolve} of {\sl Mathematica 14}. The accuracy of the numerical solution can be tested by inspecting the effect of increasing refinements of the mesh. Refining the mesh is possible by fixing the parameter {\tt MaxCellMeasure} in the function {\tt ToElementMesh}. An accuracy test can be done with reference to the region displayed in Fig.~\ref{fig1} and for the evaluation of $\sigma$, of the Nusselt number and the Brinkmann number corresponding to the {\sf T} boundary condition. By decreasing the value of {\tt MaxCellMeasure}, one attains accurate results for $\sigma$, $Nu$ and $Br$ with at least three significant figures, as shown in Table~\ref{tab1}. In order to keep the computational time reasonably small,  for each case considered, all the forthcoming results are generated with {\tt MaxCellMeasure} set to $10^{-3}$. Table~\ref{tab1} reveals that the computational region deformed with respect to the unit circle yields larger values of $\sigma$ and $|Br|$, with a smaller value of $Nu$. Indeed, the effects of a significant wall roughness generally mean a larger wall friction and a larger viscous dissipation, with respect to the case of a smooth tube wall reported in equations \eqref{24} and \eqref{25}. Furthermore, the surface roughness induces a relatively smaller velocity close to the wall resulting in a less efficient heat transfer and, hence, in a smaller $Nu$.

\begin{figure}[t]
\centering
\includegraphics[width=0.35\textheight]{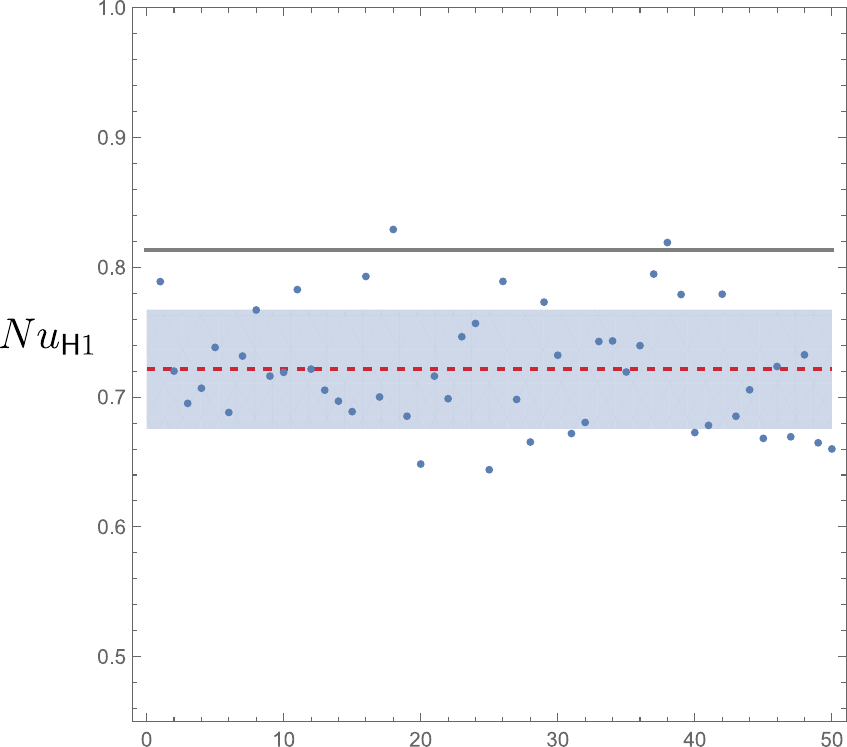}
\caption{\label{fig4}\rm\small Values of $Nu$ with the {\sf H1} boundary condition with $Br=1$ (blue dots) for the 50 different regions; the red dashed line indicates the average value $\langle Nu \rangle$ and the blue band denotes the range $\langle Nu\rangle\pm\Delta Nu$, where $\Delta Nu$ is the standard deviation; the grey line is for $Nu=48/59$ relative to the smooth circular microchannel}
\end{figure}

\begin{figure}[ht]
\centering
\includegraphics[width=0.35\textheight]{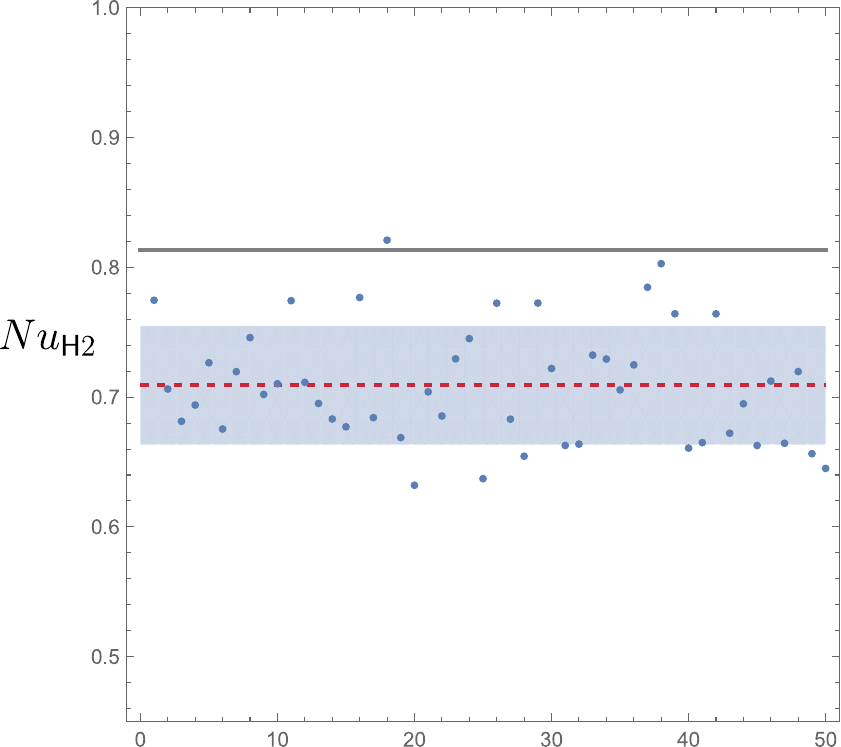}
\caption{\label{fig5}\rm\small Values of $Nu$ with the {\sf H2} boundary condition with $Br=1$ (blue dots) for the 50 different regions; the red dashed line indicates the average value $\langle Nu \rangle$ and the blue band denotes the range $\langle Nu\rangle\pm\Delta Nu$, where $\Delta Nu$ is the standard deviation; the grey line is for $Nu=48/59$ relative to the smooth circular microchannel}
\end{figure}

\section{Discussion of the results}
A statistical sample of $50$ different microchannels is obtained by generating multiple polygonal paths and, hence, multiple computational regions where the values of $\sigma$ and $Nu$, for the three cases {\sf T}, {\sf H1} and {\sf H2}, are evaluated. Such values have been reported in Figs.~\ref{fig2}-\ref{fig5} where, in abscissa, the labels of the 50 different microchannels are given. The values of $\sigma$ evaluated according to equations \eqref{24}-\eqref{26} are given for comparison as grey lines, whereas the actual average values are identified as red dashed lines with a blue coloured band showing the interval within the standard deviation tolerance. The average values and standard deviations for $\sigma$ and $Nu$, in the three cases {\sf T}, {\sf H1} and {\sf H2}, are reported in Table~\ref{tab2} where they are compared with the corresponding values obtained analytically for the smooth circular duct (second column). The last column in Table~\ref{tab2} displays the relative errors expressed as ratios between standard deviation and average value for each case. 

In all cases, it turns out that the values obtained analytically for the circular duct are more than one standard deviation away from the average values. Such a result is both evident from Figs.~\ref{fig2}-\ref{fig5} and from Table~\ref{tab2}. As a consequence, one may infer a systematic of the wall roughness, already anticipated when commenting on Table~\ref{tab1}. There is a shape sensitivity of the values of $\sigma$ and $Nu$ which results into a net increase of $\sigma$ when a surface roughness of the microchannel wall is allowed within a range of $\pm10\%$ of the duct radius. If such a roughness is unrealistic for macroscale circular ducts, it becomes quite conceivable as the size decreases to the microscale. 

The increase in $\sigma$ is due to the augmented hydraulic resistance induced by the wall roughness. A side effect is the lower velocity close to the walls caused by the meandering geometry of the perimetral path enclosing the microchannel cross-section. 

\begin{figure}[t]
\centering
\includegraphics[width=0.35\textheight]{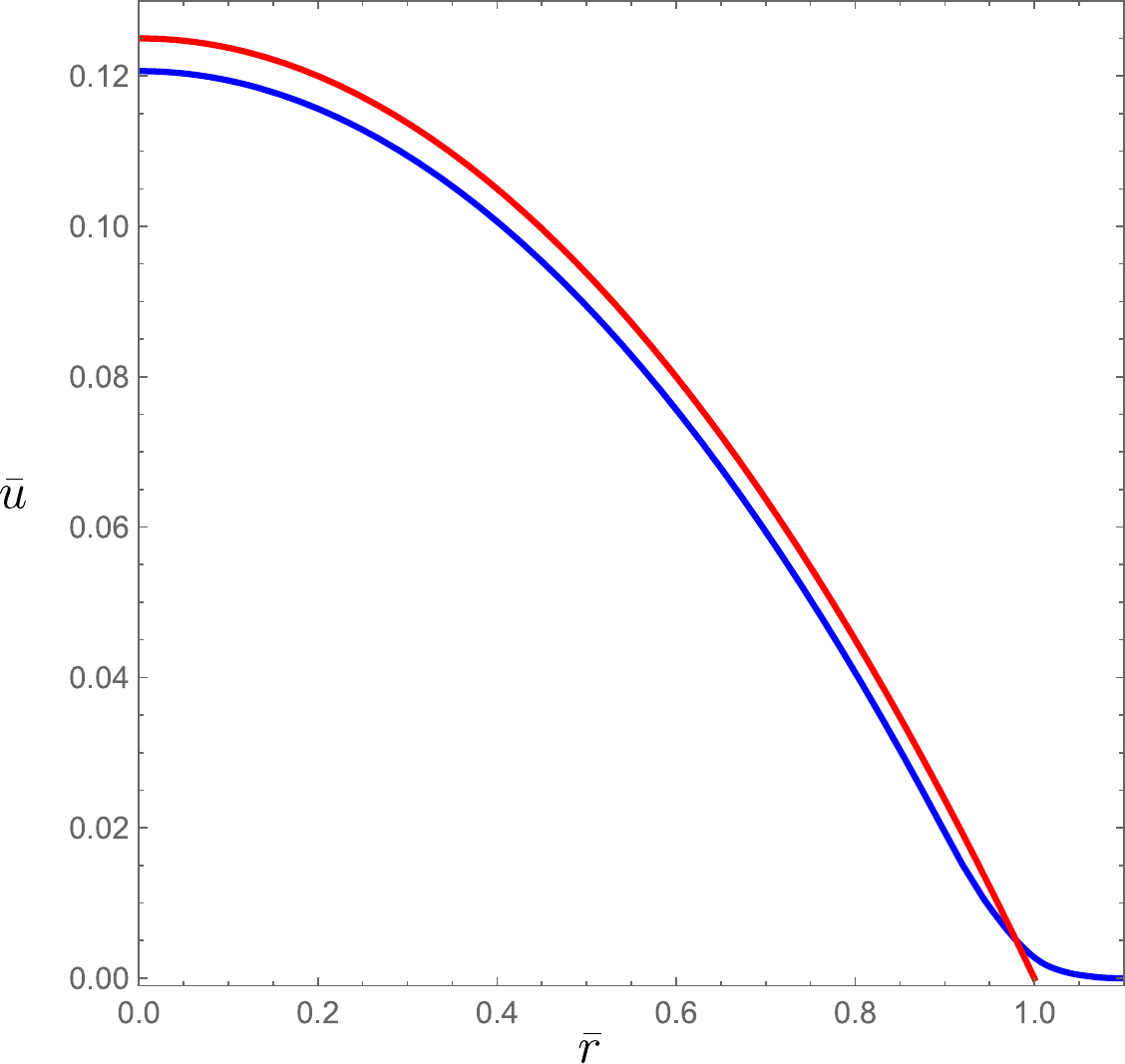}
\caption{\label{fig6}\rm\small Average radial velocity profile (in blue) obtained for the computational domain defined in Fig.\ref{fig1} compared with the Hagen-Poiseuille profile (in red) given by \eqref{22}}
\end{figure}

Fig.~\ref{fig6} illustrates the above mentioned velocity damping through a comparison between the Hagen-Poiseuille profile (red) and the average radial velocity profile (blue) obtained for the computational domain shown in Fig.~\ref{fig1}.
Such a side effect may explain the less efficient heat transfer at the wall and, hence, the lower Nusselt number with respect to the smooth circular case. In fact, lower values of $Nu$ with respect to the smooth circular duct are clearly displayed in Figs.~\ref{fig3}-\ref{fig5} relatively to the thermal boundary conditions {\sf T}, {\sf H1} and {\sf H2}, and quantitatively reported in Table~\ref{tab2}. One may mention that the velocity reduction close to the wall competes with the exchange surface increase that tends to enhance the heat transfer rate. Indeed, there are occasional instances where the latter effect prevails over the former one. This is evidenced in just two instances over the fifty considered in the {\sf H1} case (Fig.~\ref{fig4}), and in just one instance in the {\sf H2} case (Fig.~\ref{fig5}). Such exceptions are easily detected as cases where $Nu$ exceeds the smooth circular duct value. 

As noted by \citet{pelevic2016heat}, the reason behind a predicted heat transfer enhancement or reduction remains unclear. In fact, as conjectured by \citet{CROCE2004601} and by \citet{pelevic2016heat}, the heat transfer rate is expected to be unaffected by surface roughness on the average. Such a conjecture is not confirmed by the present analysis.

\section{Conclusions}
The laminar forced convection in a circular microchannel has been investigated by considering the effect of viscous dissipation and by modelling the wall roughness. 
The wall roughness has been assumed to affect the wall geometry in a random way within a range of $\pm 10\%$ of the duct radius. 

\begin{table}[t]
\centering
\begin{tabular}{|l|p{11mm}|l|p{13mm}|p{11mm}|}
\hline
quantity & circular \newline duct &  average & standard \newline deviation & relative \newline error \\
\hline
$\sigma$   & 16.00    & 17.70 & 0.54 & 3.1\,\%\\
\hline
$Nu_{\sf T}$  &  ~~9.60 & ~~7.70 & 0.27 & 3.6\,\%\\
\hline
$Nu_{\sf H1}$  &  ~~0.814 & ~~0.721 & 0.046 & 6.4\,\%\\
\hline
$Nu_{\sf H2}$  & ~~0.814 & ~~0.709 & 0.046 & 6.4\,\%\\
\hline
\end{tabular}
\caption{\label{tab2}\rm\small Data for $\sigma$ and $Nu$ relative to the three cases {\sf T}, {\sf H1} and {\sf H2}}
\end{table}

A statistical sample of fifty different microchannels has been generated as computational domains to be employed for the solution of the governing equations. In each instance of the statistical sample, all the three standard thermal boundary conditions, {\sf T}, {\sf H1} and {\sf H2}, have been considered in order to evaluate the Nusselt number $Nu$ and the Poiseuille number $\sigma$.

The comparison between the actual microchannel and the smooth circular duct has revealed a larger Poiseuille number and a smaller Nusselt number in the rough walled microchannel. 

An improvement of this analysis can be sought by including effects of wall slip that may be specially important when gas flows at low density are examined. A much more complicated aspect, yet to be developed, is the model of a geometry where the wall roughness is not only in the circumferential direction, but also in the axial direction. In fact, such a possibility of a three-dimensional wall roughness has been explored by other authors \cite{croce2005numerical, wang2007influence, pelevic2016heat, lu2020effects}. On the other hand, three-dimensional surface roughness is disregarded in the present analysis. The surface roughness model adopted here has been conceived by the tacit assumption that the technology employed for the microchannel fabrication is such that the radial changes in the wall geometry are significantly larger than the axial changes.

\bibliographystyle{elsarticle-num-names} 
\bibliography{biblio}

\end{document}